\documentclass[sigconf, nonacm]{acmart}
\AtBeginDocument{%
  \providecommand\BibTeX{{%
    \normalfont B\kern-0.5em{\scshape i\kern-0.25em b}\kern-0.8em\TeX}}}

\usepackage{amsmath,amssymb,amsfonts}
\usepackage{algorithmic}
\usepackage{graphicx}
\usepackage{textcomp}
\usepackage{subcaption}
\usepackage{xcolor}
\usepackage{multirow}
\usepackage{siunitx}

\renewcommand\footnotetextcopyrightpermission[1]{}
\begin{document}

% \title{Compute-in-Memory Accelerators for Space: \\ Thermal-Resilient Alternatives to GPU and HBM}
\title{Space-CIM: Enabling Compute-In-Memory Accelerators for Thermally-Constrained Space Platforms}
\titlenote{Author preprint. Accepted to the ACM/IEEE International Symposium on Low Power Electronics and Design (ISLPED '26), August 05--07, 2026, Evanston, IL, USA. The Version of Record is available in the ACM Digital Library. DOI: 10.1145/3816440.3818579.}

%% Space-CIM:
%%Compute-in-Memory for Space: Thermal-Resilient Alternatives to GPU and HBM
% \author{Placeholder for Authors}

\author{Sohan Salahuddin Mugdho}
\orcid{0009-0004-2670-285X}
\affiliation{%
  \department{Electrical and Computer Engineering}
  \institution{Iowa State University of Science and Technology}
  \city{Ames}
  \state{IA}
  \postcode{50010}
  \country{USA}
}

\author{Md. Shahedul Hasan}
\orcid{0000-0002-7182-0677}
\affiliation{%
  \department{Electrical and Computer Engineering}
  \institution{Iowa State University of Science and Technology}
  \city{Ames}
  \state{IA}
  \postcode{50010}
  \country{USA}
}
\author{Cheng Wang}
\authornote{Corresponding author: chengw@iastate.edu}
\orcid{0000-0002-8815-2871}
\affiliation{%
  \department{Electrical and Computer Engineering}
  \institution{Iowa State University of Science and Technology}
  \city{Ames}
  \state{IA}
  \postcode{50010}
  \country{USA}
}
\begin{abstract}
The rapid growth in compute demand from artificial intelligence (AI) has driven a massive surge in data center construction, precipitating an energy and sustainability crisis. Motivated by the abundant solar energy in outer space and the recent sharp reduction in space launch costs, orbital data centers are emerging as a potential pathway for the future scaling of AI compute infrastructure. While the cold background in vacuum seems appealing for cooling, computing systems operating in space without convection ultimately rely on radiative cooling, requiring large-area radiators. Such limitations in thermal management pose a significant challenge for deploying the standard liquid/air-cooled computers in space. In this work, we investigate the impact of the thermal constraints in space on both graphics processing units (GPUs) with high-bandwidth memory (HBM) and the emerging compute-in-memory (CIM) accelerators. We develop a radiator-in-the-loop co-design methodology that directly links the permitted system TOPS (terra-operations per second) with the practical radiator cooling capacity in space. Our thermal simulations reveal that the separately located GPU die and HBMs create severe thermal hotspots under limited radiator capacity, necessitating GPU thermal throttling. In contrast, CIM accelerators exhibit a much more uniform heat distribution and consistently outperform GPUs in TOPS/W across a wide range of radiator budgets. We systematically evaluated the performance of CIM and GPU across various AI workloads and demonstrated that CIM has a magnified advantage for deployment in space under realistic thermal constraints.
\end{abstract}
%%\vspace{-10pt}



% \keywords{AI Accelerators, Compute-In-Memory, Radiative Cooling}



\maketitle

\section{Introduction}
%\vspace{-2pt}
Artificial Intelligence (AI) based on Deep Neural Networks (DNNs) has achieved remarkable cognitive capability across diverse domains, including language processing, computer vision, autonomous driving, and healthcare \cite{deeplearning, resnet,autonomous-driving, healthcare}. These advances are accompanied by an exponential growth in compute complexity and input/model size, imposing an ever-increasing demand for heavy computing resources. At present, graphics processing units (GPUs) with high-bandwidth memory (HBM) are the major workhorses fueling AI computation, thanks to their massive processing parallelism \cite{ml-workload} combined with specialized high memory bandwidth. However, building large-scale AI data centers based on the power-hungry GPU+HBM platforms has created an immense challenge for the underlying energy infrastructure. It has been projected that data center workload will make up 30-40 percent of all new energy demand in the United States by 2030 \cite{datacenters_power_2024}.
Therefore, the rapid growth of data center energy consumption has become a societal and environmental concern, motivating the exploration of the unconventional deployment of AI data centers in environments where abundant energy can be harnessed at scale.
\begin{figure*}[hbt]
    \centering
    \includegraphics[width =0.95\linewidth]{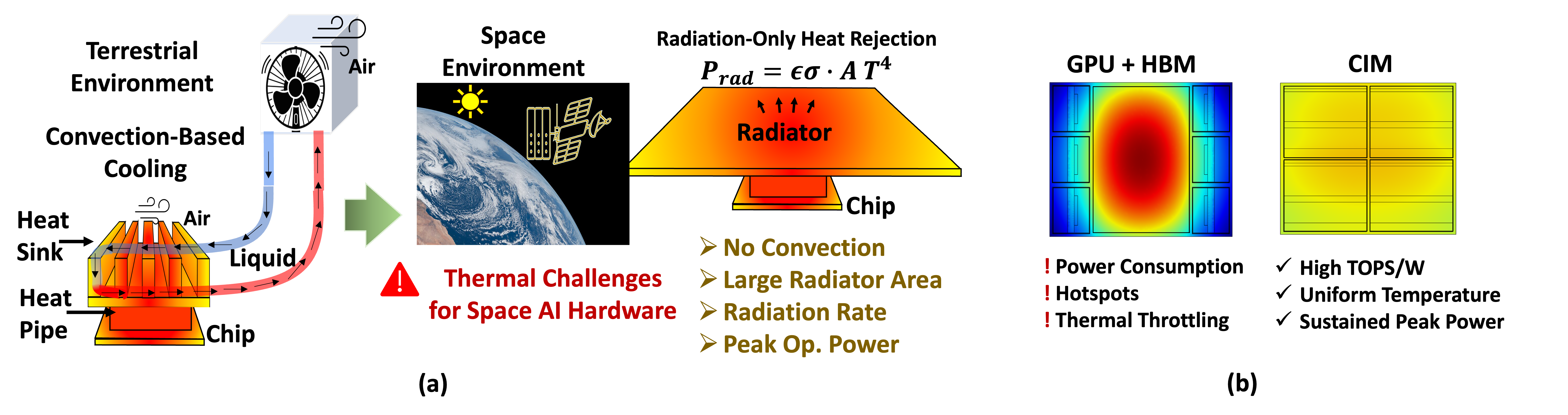}
    \vspace{-10pt}
    \caption{(a) An overview of the key thermal management challenges for space AI hardware compared to terrestrial hardware. (b) Illustration of the temperature profiles of the GPU and the CIM chips under limited cooling.}
    \vspace{-5pt}
    \label{fig:overview}
\end{figure*}

Space (orbital) data centers have recently emerged as a promising solution to the future scaling of AI compute infrastructure, thanks to the immense potential of the continuous and high-intensity solar energy source and virtually unlimited physical room for large-scale deployments \cite{space-datacenter-1}. However, while the cold background temperature in space may seem attractive for keeping the compute chips cool, AI platforms working in the space vacuum must operate without convection. As a result, cooling becomes a critical challenge for space computing infrastructure as chips must \textbf{rely ultimately on thermal radiation} for heat exchange with the surrounding \cite{space-radiator-1}. Unlike convection-based cooling on earth, where small heat sinks can reject substantial heat, radiative cooling demands large and heavy radiators, severely constraining the operating power of the supported AI compute systems. The volume and mass of radiators now become crucial for designing space computing systems. 

Standard terrestrial GPUs typically rely on a combination of liquid and fan cooling to reach near-peak performance while maintaining a safe working temperature \cite{a100_datasheet}. Hence, if deployed in space, the severe thermal constraints will inevitably impact the computational performance of standard GPU+HBM systems.
It is important to note that the majority of the energy consumption with GPUs is attributed to the data movement between HBM and the GPU compute logic \cite{hbm-bw-power-data}, commonly known as the von Neumann memory wall bottleneck. 
To tackle the issues of heat generation and concentration in GPU+HBM systems, compute-in-memory (CIM) architecture with various memory technologies has been extensively investigated to seek an alternative pathway to compute efficiency\cite{memory-table}. In particular, CIM accelerators based on non-volatile memory (NVM) offer high on-chip memory density and low static power, suggesting immense potential for inference acceleration.
While most research on CIM focuses on demonstrating energy-efficiency improvements for AI computation, it remains largely unexplored how these CIM systems perform under passive thermal constraints (such as given a radiator rejection power budget). 

In this work, we systematically investigate how CIM accelerators perform in comparison with standard GPU+HBM under realistic cooling constraints bounded by the availability of radiators.
We demonstrate that crossbar-based CIM architectures mitigate the memory bottleneck and manage to both (1) achieve improved power efficiency and (2) eliminate the thermal hotspots. To the best of our knowledge, \textbf{our work is the first design exploration of CIM accelerators for space computing with radiator constraints} in the loop. The major contributions of the paper are the following:
% %\vspace{-10pt}
\begin{itemize}
    \item We conduct finite-element-methods (FEM)-based system modeling to assess the impact of limited cooling on AI hardware in space. Under the iso-chip-area condition with fixed radiator thermal rejection power, the compute-efficient CIM results in lower overall chip temperature. Interestingly, compared to GPU+HBM systems that exhibit strong thermal hotspots, CIM accelerators exhibit a homogeneous spatial thermal profile, thereby eliminating the hotspot issue.

    \item A radiator-in-the-loop thermal-electronic co-design methodology is developed for optimizing CIM architecture for thermally constrained space platforms. Concretely, we propose an evaluation framework based on a roofline-like modeling that explicitly links radiator availability to the impact of thermal throttling on peak performance. The performance improvement of CIM over standard GPU systems is quantified based on the proposed modeling framework.
   
    \item Various CIM architectural configurations are evaluated for a diverse group of representative AI compute workloads. Our system evaluation based on the thermal-aware roofline modeling demonstrates that CIM accelerators consistently deliver over 10-40x improvement in TOPS compared to GPU+HBM under tight thermal constraints. The magnified performance advantage of CIM over GPU opens up an exciting venue for developing AI hardware for space computing.
\end{itemize}
%\vspace{-10pt}
\section{Related Works}
\subsection{Thermal Management for AI Hardware}
Cooling is a key design concern for large-scale computing infrastructure. Advanced thermal management packaging cools chips in terrestrial environments, and liquid and air cooling are combined in state-of-the-art data centers\cite{understand_cooling,intro-ack-1}. Beyond rapid heat removal, thermal management must address thermal coupling between on-chip components under non-uniform heat generation. In conventional GPU+HBM systems (e.g., GPU A100), most power is concentrated on the GPU die, creating high-power-density hotspots\cite{hbmgpu_thermal_ref_paper}. Thermal coupling between the GPU and temperature-sensitive HBM dies triggers throttling and lowers effective throughput below peak\cite{gpu-capping}. Dynamic Voltage and Frequency Scaling (DVFS), the standard GPU power management technique, balances performance, power, and heat as a reactive safeguard against overheating\cite{gpu-dvfs-ref-2, gpuwattch}. As detailed later, our analysis shows that both total heat and non-uniform spatial thermal profiles are harder to manage in space, causing significant performance degradation for standard GPUs on space platforms.
\vspace{-10pt}
\subsection{Space AI Compute}
Space AI compute is drawing growing academic and industrial interest, enabled by continuous solar power without weather losses and the possibility of passive radiative cooling to vacuum. As technologies mature, space compute will first support onboard processing of satellite and orbital sensor data (e.g., remote sensing, telescope imaging), then progress toward full data-center-class AI compute in low Earth orbit (LEO). Though nascent, the field has already achieved milestones, including the first in-space use of an H100 GPU \cite{space-datacenter-2} to run intermediate-size AI workloads such as nanoGPT training \cite{nanoGPT}. 

Major challenges remain for large-scale deployment, notably thermal management without convection and ensuring electronics reliability under intense cosmic radiation. Radiation-hardening techniques from space and mission-critical electronics can protect AI hardware \cite{radiation_harden}, and recent radiation-tolerance tests on Google TPUs show promising accelerator resilience \cite{suncatcher_TPU}. Additional challenges span the deployment life cycle, including launch costs, SWaP (Size, Weight, and Power) constraints, inter-satellite and satellite– ground communication, and maintenance logistics. This work focuses on the thermal constraint and proposes a hardware solution.
\begin{table}
\centering
\caption{Physical dimension and power breakdown of GPU and CIM systems for thermal modeling.}
%\vspace{-12pt}
\label{tab:gpu_cim_breakdown}
\begin{tabular}{lccc}
\toprule
\multicolumn{4}{c}{\textbf{(a) GPU (based on \cite{hbmgpu_thermal_ref_paper})}} \\
\midrule
\textbf{Component} & \textbf{Dimensions} & \textbf{Power} & \textbf{Count} \\
                   & \textbf{(W$\times$H$\times$t) (mm)} & \textbf{(W)} & \\
\midrule
Interposer   & 54$\times$50$\times$0.11 & NA        & 1 \\
GPU-Core     & 34.4$\times$28$\times$0.72 & 20--320  & 1 \\
Logic        & 11$\times$11$\times$0.1   & NA        & 1 \\
Logic-TSV    & 2$\times$10.6$\times$0.1  & 1.22      & 1 \\
Logic-PHY    & 1$\times$7$\times$0.1     & 2.44      & 1 \\
DRAM-Bank    & 10.6$\times$10.6$\times$0.77 & 0.92  & 8 \\
DRAM-TSV     & 2$\times$10.6$\times$0.77 & 0.3       & 8 \\
HBM-stack    & 11$\times$11$\times$0.72  & 13.4      & 6 \\
\midrule
\multicolumn{4}{c}{\textbf{(b) CIM (based on CiMLoop \cite{cimloop})}} \\
\midrule
\textbf{Component} & \textbf{Dimensions} & \textbf{Power} & \textbf{Count} \\
                   & \textbf{(W$\times$H$\times$t) (mm)} & \textbf{(W)} & \\
\midrule
Interposer   & 52$\times$52$\times$0.1  & NA        & 1 \\
CIM-Chip     & 25$\times$25$\times$0.72 & 16.25--60 & 4 \\
CIM-Unit     & 25$\times$0.96$\times$0.72 & 2.08--7.67 & 1 \\
Row-Drivers  & 25$\times$1.04$\times$0.72 & 0.89--3.28 & 1 \\
ADC          & 25$\times$10.44$\times$0.72 & 7.92--29.22 & 1 \\
Digital      & 25$\times$2.2$\times$0.72 & 0.93--3.43 & 1 \\
NoC          & 25$\times$3.1$\times$0.72 & 4.19--15.46 & 1 \\
Buffer       & 25$\times$7.26$\times$0.72 & 0.21--0.77 & 1 \\
\bottomrule
\vspace{-15pt}
\end{tabular}
\end{table}
%\vspace{-5pt}

\subsection{Compute-In-Memory Accelerators}
% %\vspace{-5pt}
To overcome the von Neumann memory bottleneck in standard digital hardware, memory-centric architectures bring memory and compute units together (“in-memory”) or near each other (“near-memory”) \cite{memory-wall}. Crossbar-based Compute-In-Memory (CIM) with analog accumulation is a promising AI acceleration paradigm: it performs MAC operations within memory cells and uses Kirchhoff's law across columns for analog accumulation, enabling massively parallel, \textit{in-situ} matrix-vector multiplications (MVM) with greatly reduced data movement \cite{memory-table}. Experiments show that crossbar CIM can effectively mitigate the memory wall and potentially surpass GPU+HBM systems in energy efficiency, measured in terms of Trillion Operations Per Second per Watt (TOPS/W) \cite{intro-ack-5,hermes}. CIM also yields a more uniform spatial heat distribution by spreading matrix-vector workloads across memory arrays \cite{issac}, eliminating hotspots seen in HBM+GPU systems. As detailed later, our evaluation shows that CIM systems not only generate less total heat due to higher energy efficiency, but also alleviate thermal hotspots under tight power and thermal constraints.
\begin{figure}
    \centering
    \includegraphics[width=1.03\linewidth]{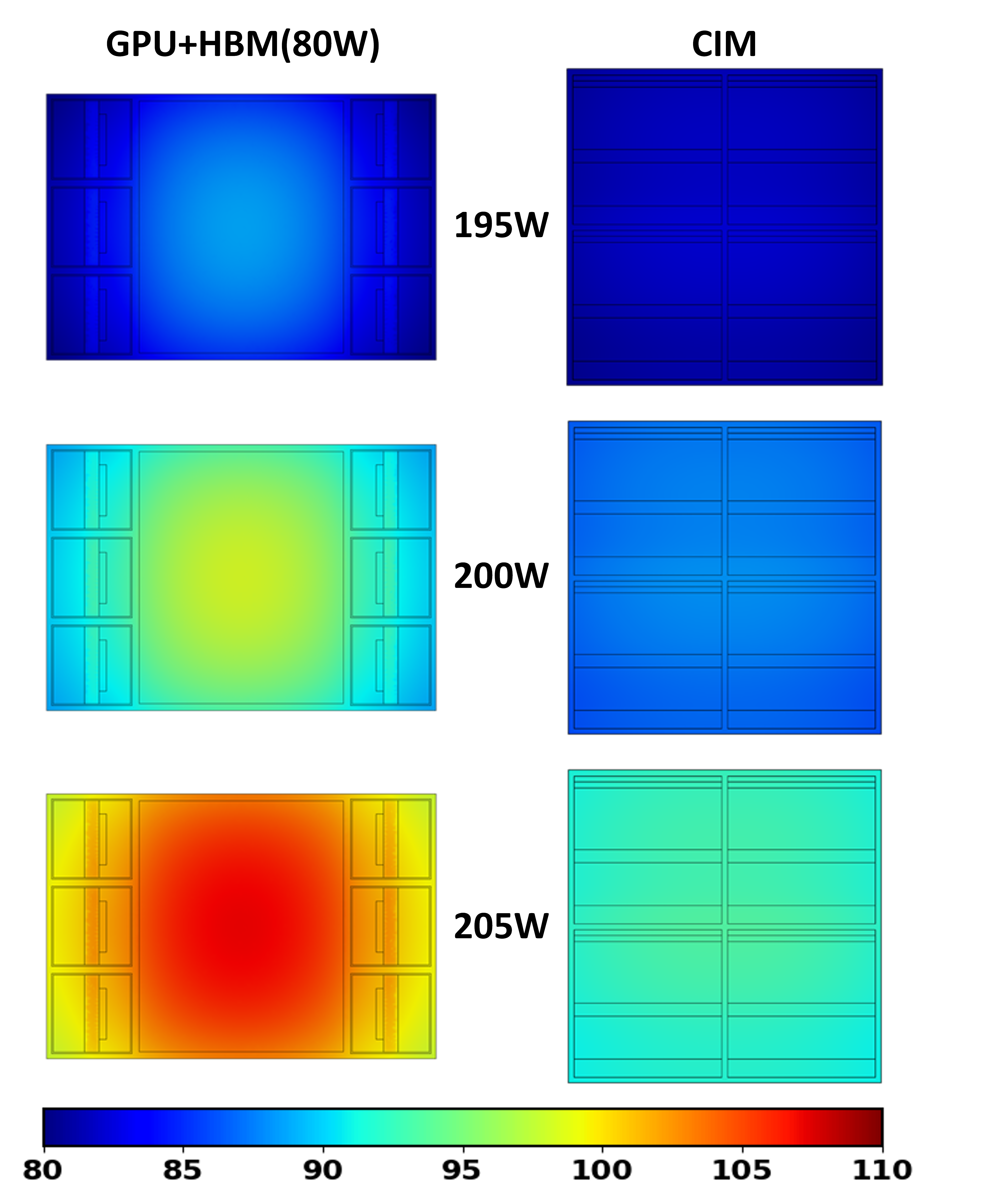}
    \vspace{-10pt}
    \caption{Thermal maps of GPU+HBM (top) and CIM (bottom) systems at different total power levels relative to a Thermal Rejection Power (TRP) of 200 W @ \SI{85}{\celsius}, highlighting the growing hotspot at the GPU core.}
    \vspace{-15pt}
    \label{fig:gpu-cim-heatmap}
\end{figure}
% %\vspace{-10pt}
\section{Thermal Characterization of AI accelerators}
\label{sec:thermal}
\textbf{Thermal modeling setup.} We characterize the thermal behavior of GPU+HBM and CIM-based AI accelerators under varying operating and cooling conditions. Using Comsol \cite{comsol51}, we model GPU, HBM, and CIM package components and run finite element method (FEM) solid heat-transfer simulations. The goal is to accurately evaluate temperature distribution under different power levels and radiator thermal rejection power (TRP) budgets. The GPU core is modeled as a monolithic silicon (Si) die. The HBM is modeled stack-wise, including the physical I/O on the logic die (Logic-PHY), through-silicon vias (TSVs) connecting the logic layer (Logic-TSV) and DRAM layers (DRAM-TSV), and DRAM banks, following \cite{hbmgpu_thermal_ref_paper}. HBM dimensions and power distribution are scaled to an 8-Hi stack configuration with six stacks at a combined 80 W and 1.9 TB/s bandwidth \cite{hbm-bw-power-data}, matching an NVIDIA A100-class GPU+HBM system at 400 W and 624 TOPS \cite{a100_datasheet}. The full GPU+HBM package is placed on a silicon interposer with Si dies for DRAM and Logic-PHY and $SiO_2$-encased Cu TSVs. The CIM system is modeled as four Si chips on a silicon interposer with a form factor similar to the A100-like GPU+HBM system. Component area and power are estimated using CiMLoop \cite{cimloop} with parameters from \cite{issac}, summarized in Table~\ref{tab:gpu_cim_breakdown}. Memory cell/array configurations and the peripheral circuits for analog-digital converters (ADC) will be varied for CIM design-space exploration. The radiator is modeled as a flat panel rejecting $100~W/mm^2$ \cite{rad-area-power-ref} at \SI{85}{\celsius}.

\textbf{Comparison of thermal maps.} 
Fig.~\ref{fig:gpu-cim-heatmap} shows thermal maps for GPU+HBM and a representative CIM system at 195 W, 200 W, and 205 W under a fixed 200 W TRP. In GPU+HBM systems, the GPU core has the highest power density and dominates heat generation, quickly reaching high temperatures even within the TRP budget, while thermal coupling raises temperatures in the thermally sensitive HBM stacks. In contrast, CIM systems avoid high power-density hotspots, yielding a more uniform thermal profile and safe operation within TRP limits.
\begin{figure}
    \centering
    %\vspace{-5pt}
    \includegraphics[width=1.00\linewidth]{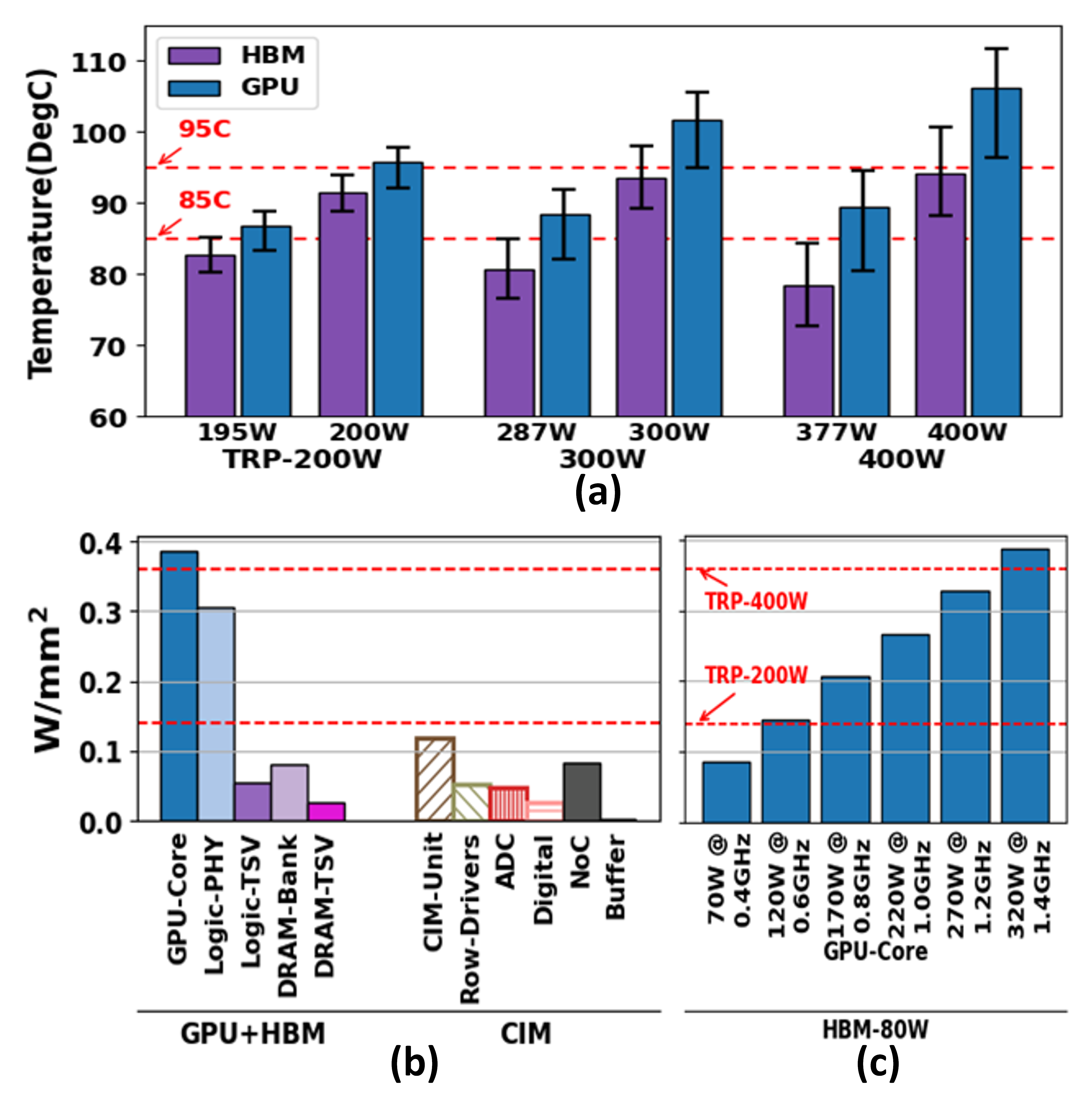}
    \vspace{-20pt}
    \caption{Thermal behavior of GPU+HBM and CIM under TRP constraints (@ \SI{85}{\celsius}). (a) shows operation at full TRP and power scaling to maintain safe GPU (\SI{95}{\celsius}) and HBM (\SI{85}{\celsius}) temperatures, (b) compares component-wise thermal density, and (c) illustrates GPU-core power–frequency scaling to meet safe operation under different TRPs.}
    \vspace{-15pt}
    \label{fig:temperature-profile}
\end{figure}
% %\vspace{-10pt}
The GPU+HBM system consistently shows a severe hotspot at the GPU core. While total power stays within the TRP limit, hotspot severity increases with power, indicated by a growing temperature gap across chip regions. In contrast, the CIM system sustains lower average temperatures and an almost uniform thermal profile; simulations show that CIM eliminates hotspots even at higher power.

\textbf{Quantification of elevated temperature.} High power density and thermal coupling push both the GPU core and HBM above safe temperatures under nominal TRP. The GPU core must throttle when the logic die exceeds \SI{95}{\celsius}. Elevated HBM temperatures increase refresh rates, cutting effective bandwidth to about 73\% at \SI{85}{\celsius} \cite{hbmgpu-thermal}. Fig.~\ref{fig:temperature-profile}(a) shows that, even at the limited TRP, GPU core and HBM exceed safe limits, requiring system-level power reduction to maintain reliability.

\textbf{Power density analysis.} We analyze power density across GPU+HBM and CIM components. The GPU+HBM system includes high-density regions such as the GPU-core and Logic-PHY, while the CIM system comprises RRAM crossbar arrays, row drivers, ADCs, digital units, NoC (network on chip interconnects), and buffers. As shown in Fig.~\ref{fig:temperature-profile}(b), the GPU-core and Logic-PHY have much higher power densities than other components, whereas CIM components remain low and nearly uniform. Fig.~\ref{fig:temperature-profile}(b) also shows GPU-core power density under different DVFS operating points, along with two TRP capacity levels indicating power density headroom. The GPU-core must greatly reduce its operating frequency to operate safely within TRP constraints.
\begin{figure}
    \centering
    \includegraphics[width=1.03\linewidth]{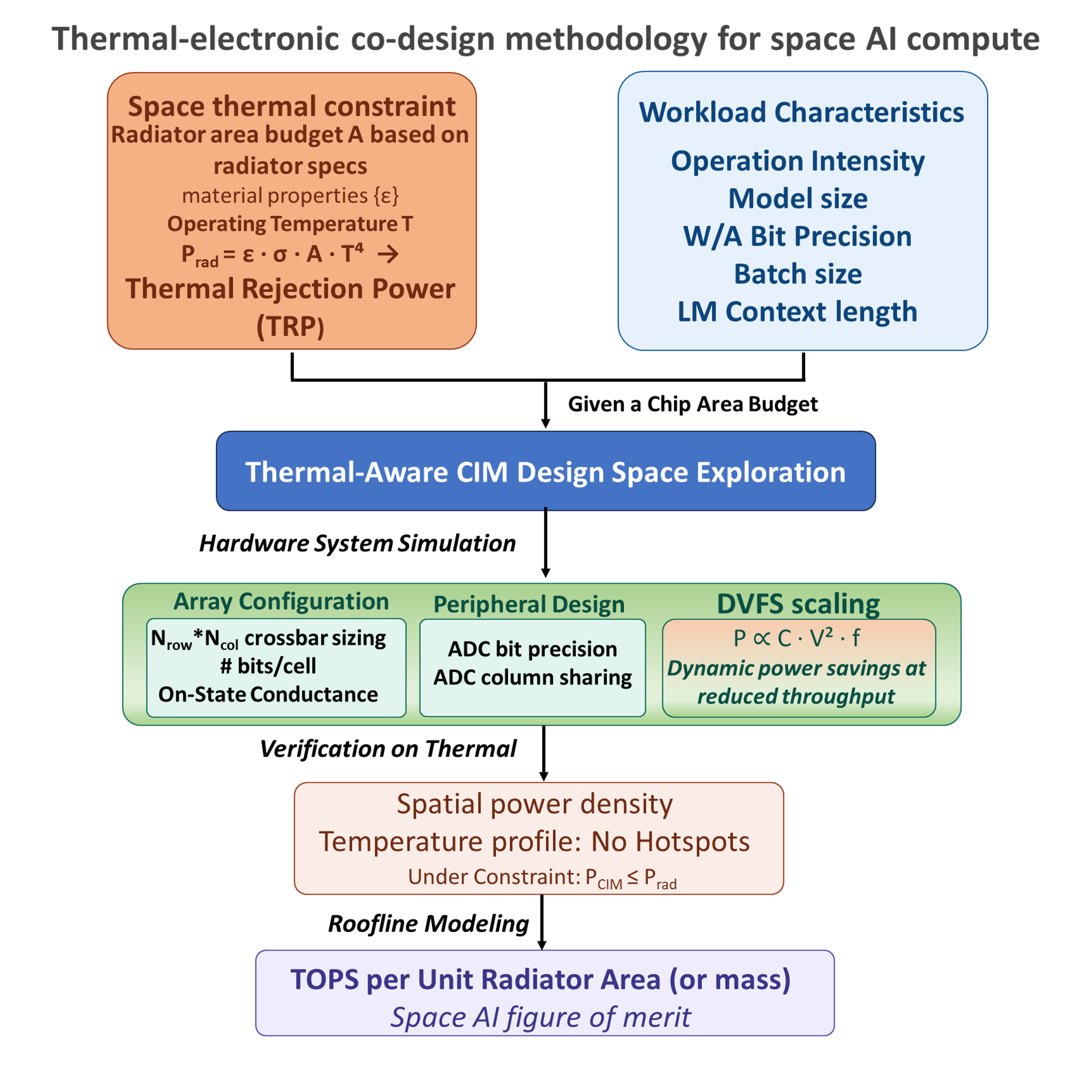}
    \vspace{-20pt}
    \caption{Overview of the proposed thermal-electronic co-design workflow.}
    \vspace{-10pt}
    \label{fig:cim-workflow}
\end{figure}
Particularly, more than 50\% clock frequency reduction is needed to accommodate a TRP of 200 W.
\section{Evaluation Methodology}
\label{sec:evaluation}
% \textbf{Space thermal constraints.} 
We develop a thermal-aware hardware simulation framework to evaluate the performance of GPU+HBM and different CIM configurations under space thermal constraints. In space environments, AI accelerators rely entirely on radiative cooling for safe operation. Based on the Stefan-Boltzmann law, $P_{rad} = \epsilon \cdot \sigma \cdot A \cdot (T^4 -T_{ext}^4) $, the radiator area and material emissivity $\epsilon$ determine the maximum thermal power that can be dissipated at a given operating temperature ($T$) and external temperature ($T_{ext}$). The external temperature $T_{ext}$ is set at
$140~K$ as a low end for the temperature range of near-earth orbits \cite{LEO_link}, which is negligible compared to the operating temperature of \SI{85}{\celsius}. 
For our experiments, we conservatively assume a radiator thermal rejection capability of 100 $W/m^2$ \cite{rad-area-power-ref}, making radiator area and TRP the primary thermal constraints.

\textbf{DVFS performance scaling.} To operate GPU+HBM and CIM configurations under different TRP (or radiator area) constraints, dynamic voltage-frequency scaling (DVFS) is applied. Under DVFS, system power scales with voltage ($V$) and frequency ($f$) as $P \propto C \cdot V^2 \cdot f$, which can be approximated as $P \propto f^\alpha$ based on the $V$–$f$ relationship. In modern GPUs, voltage remains approximately constant at low frequencies and scales linearly with frequency at higher operating points \cite{gpu-dvfs-ref-1, gpu-dvfs-ref-2}. Using empirical power–frequency data from \cite{gpu-dvfs-ref-1}, we model this behavior with $\alpha = 1$ for $f < f_{breakpoint}$ and $\alpha = 2.1$ for $f \geq f_{breakpoint}$.

\textbf{Thermal-aware CIM design.} The DVFS model for CIM systems follows a simpler relationship due to the presence of mixed-signal peripheral components. We assume a fixed operating voltage across the frequency range, resulting in $P \propto f$. Based on these DVFS models, we estimate the achievable performance (TOPS) under given TRP constraints. Fig.~\ref{fig:cim-workflow} illustrates the workflow for thermally-aware 
\begin{table}
\centering
\caption{System and device parameters used in evaluation.\\(*) marked values are based on CiMLoop \cite{cimloop} simulation.}
%\vspace{-10pt}
\label{tab:system_device}
\begin{tabular}{lcc}
\toprule
\multicolumn{3}{c}{\textbf{(a) System Parameters}} \\
\midrule
\textbf{System} & \textbf{GPU} & \textbf{CIM} \\
\midrule
Model & A100 \cite{a100_datasheet}& Based on \cite{issac}\\
Max Frequency (MHz) & 1410 & 20 \\
Max Power (W) & 400 & 160-240*\\
Min Power (W) & 100 & 65-100*\\
Static Power (\%) & 15 & 5 \\
Peak TOPS & 624 & 2658-5755*\\
Op. Bits & 8 & 8 \\
Memory & HBM & DDR5 \\
Memory Power (W) & 80 \cite{hbm-bw-power-data}& 60 \cite{hbm2}\\
Bandwidth (GB/s) & 1935 \cite{hbm-bw-power-data}& 536 \cite{hbm2}\\
Technology (nm) & 7 & 16 \\
\midrule
\multicolumn{3}{c}{\textbf{(b) Device Parameters}} \\
\midrule
\textbf{Property} & \textbf{RRAM} \cite{puma}& \textbf{STT-MRAM} \\
\midrule
Cell Size ($F^2$) & 27 & 73 \cite{mram-area-ref}\\
$R_{\mathrm{on}}$ & 100 & 13 \cite{mram-ron-roff}\\
$R_{\mathrm{off}}/R_{\mathrm{on}}$ & 10 & 2 \cite{mram-ron-roff}\\
Bits/Cell & 1, 2 & 1 \\
\bottomrule
%\vspace{-20pt}
\end{tabular}
\end{table}
CIM design space exploration, considering varying thermal constraints, workload characteristics, and hardware parameters such as crossbar array configurations and peripheral designs. As established in Sec.~\ref{sec:thermal}, CIM systems avoid thermal hotspots under power constraints, enabling efficient exploration of achievable TOPS within the radiator-limited design space.
\section{Results and Discussion}
\label{sec:results}
\textbf{Experimental setup.} We perform thermal-aware CIM design space exploration based on an ISAAC-like \cite{issac} CIM architecture with multiple configurations under varying space radiator thermal rejection power limits. The CIM systems are simulated using CiMLoop, a fast design-space exploration framework for CIM architectures \cite{cimloop}, targeting a 16nm technology node. All comparisons are conducted under iso-area constraints. Specifically, the CIM system consists of four chips, each of 625 $mm^2$, integrated on a silicon interposer with a total form factor of ~2700 $mm^2$.

We compare these configurations against an NVIDIA A100 GPU+ HBM system, labeled as \textbf{GPU}. The A100 operates at a peak power of 400 W at 1.4 GHz ($f_{max-GPU}$) and a minimum power of 100 W at 165 MHz ($f_{min-GPU}$), following the DVFS model described in Sec.~\ref{sec:evaluation}. Based on empirical observations \cite{gpu-dvfs-ref-1}, the breakpoint frequency ($f_{breakpoint}$) is estimated as 1.1 GHz. The HBM subsystem operates at 80 W and delivers 1.9 TB/s bandwidth \cite{hbm-bw-power-data}, while the GPU achieves 624 TOPS for 8-bit operations \cite{a100_datasheet}. As for the static power consumption, approximately 60 W (about 15\% of peak power) is for a standard 400W GPU, while non-volatile-memory-based CIM systems have a reduced static power percentage at 5\%. 
The CIM configurations operate over a frequency range of 8 MHz ($f_{min-CIM}$) to 20 MHz ($f_{max-CIM}$), accounting for the peripheral ADC and the delay of the resistive crossbar array. 
Additionally, DDR5 DRAM is considered as off-chip memory for CIM, operating at 60 W and providing 
\begin{figure}
    \centering
    % %\vspace{-10pt}
    \includegraphics[width=0.90\linewidth]{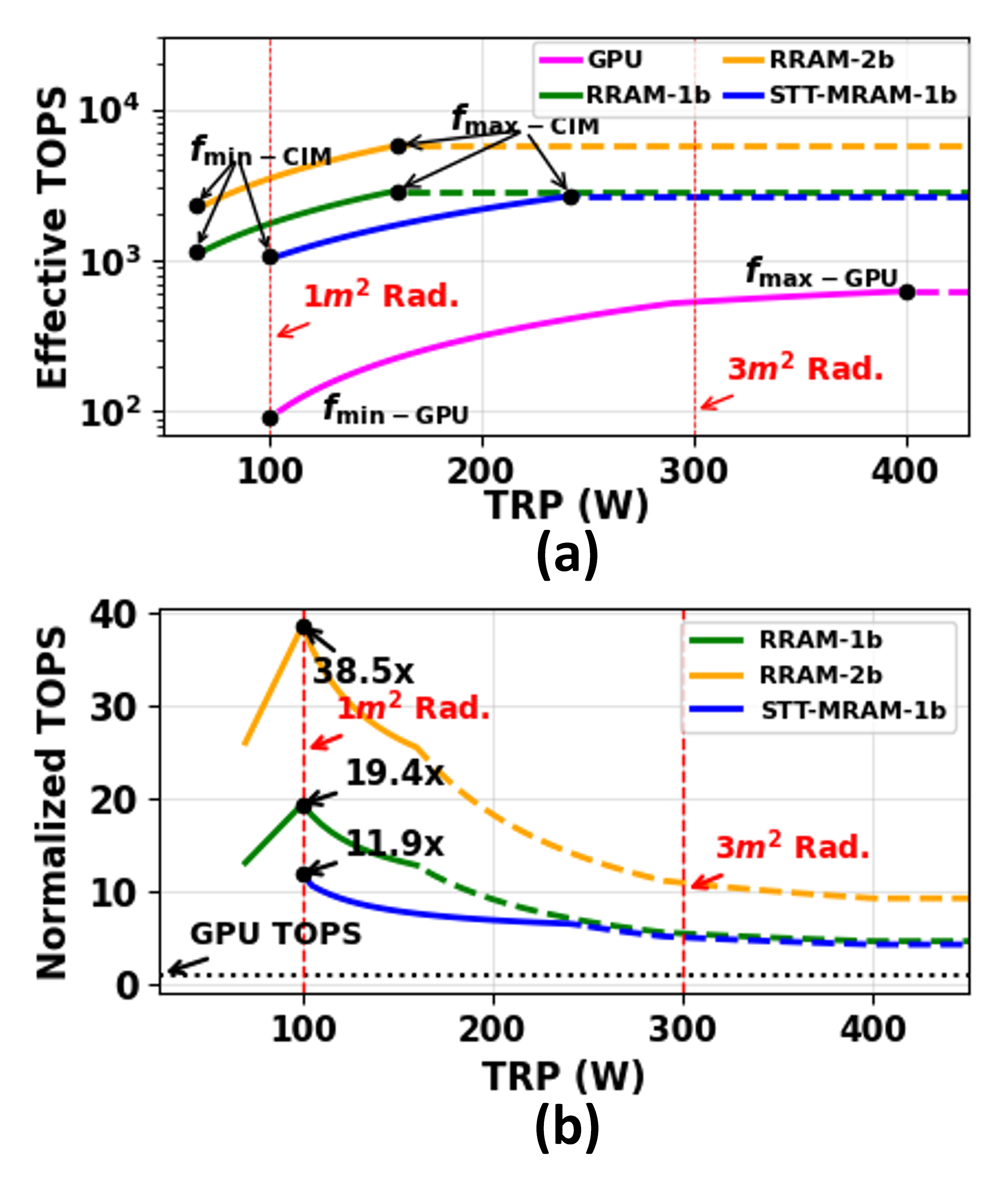}
    \vspace{-15pt}
    \caption{Performance scaling of the GPU and representative CIM systems under varying TRP budgets (@ \SI{85}{\celsius}). (a) Effective TOPS across platforms, and (b) normalized TOPS highlighting the performance advantage of CIM over GPU+HBM under TRP constraints.}
     \vspace{-10pt}
    \label{fig:cim-tops-compare}
\end{figure}
\begin{figure}
    \centering
    %\vspace{-5pt}
    \includegraphics[width=0.99\linewidth]{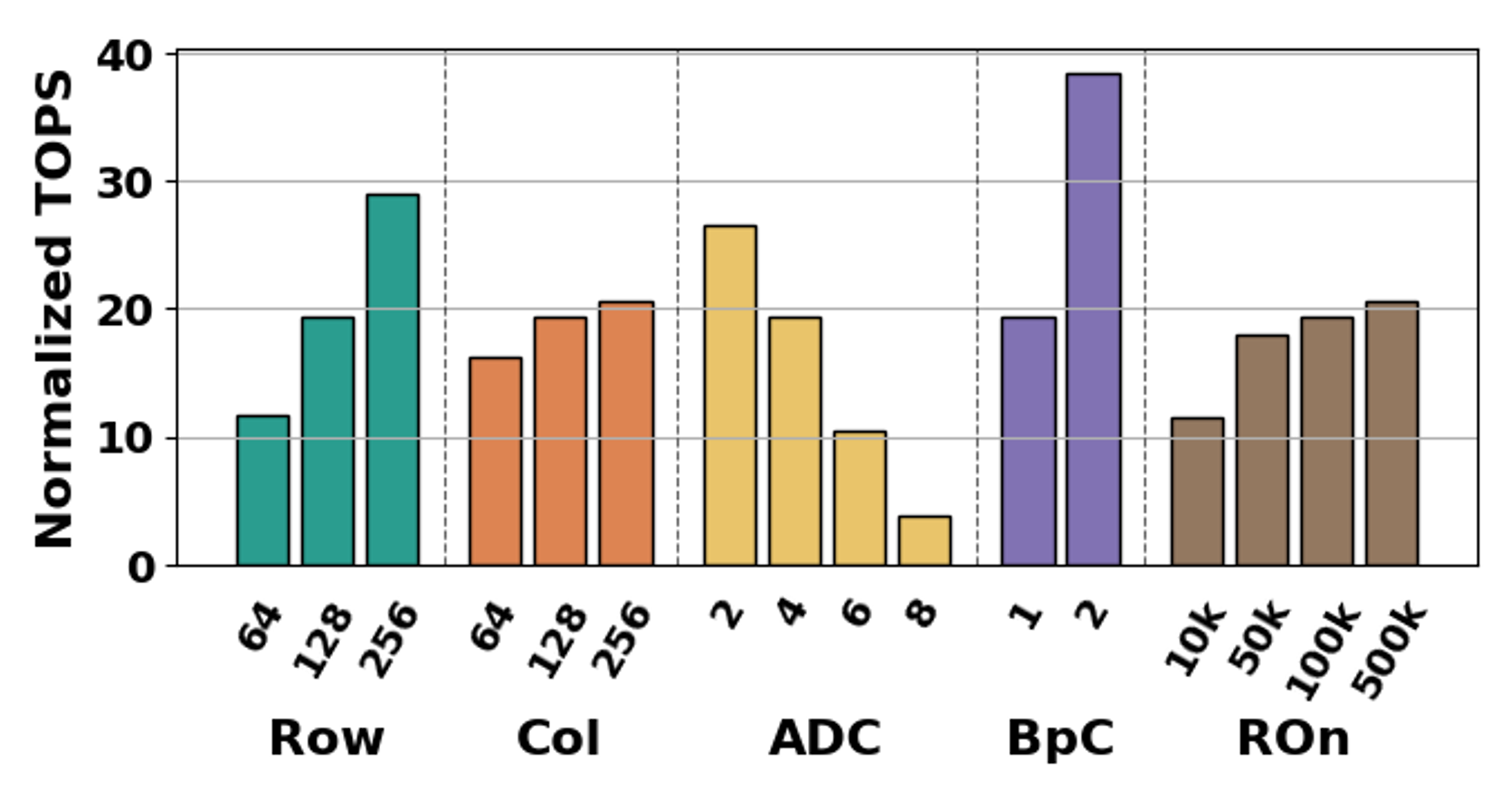}
    \vspace{-10pt}
    \caption{Design space exploration of CIM under a low TRP budget (100 W), evaluating the impact of array-level parameters on performance. Normalized TOPS (vs. GPU+HBM) highlights sensitivity to Row, Column, ADC bits, bits per crossbar cell (BpC), and cell on resistance (R\textsubscript{On}) configurations.}
    \vspace{-10pt}
    \label{fig:cim-design-space}
\end{figure}
\begin{figure*}[!t]
    \centering
    \includegraphics[width=0.95\linewidth]{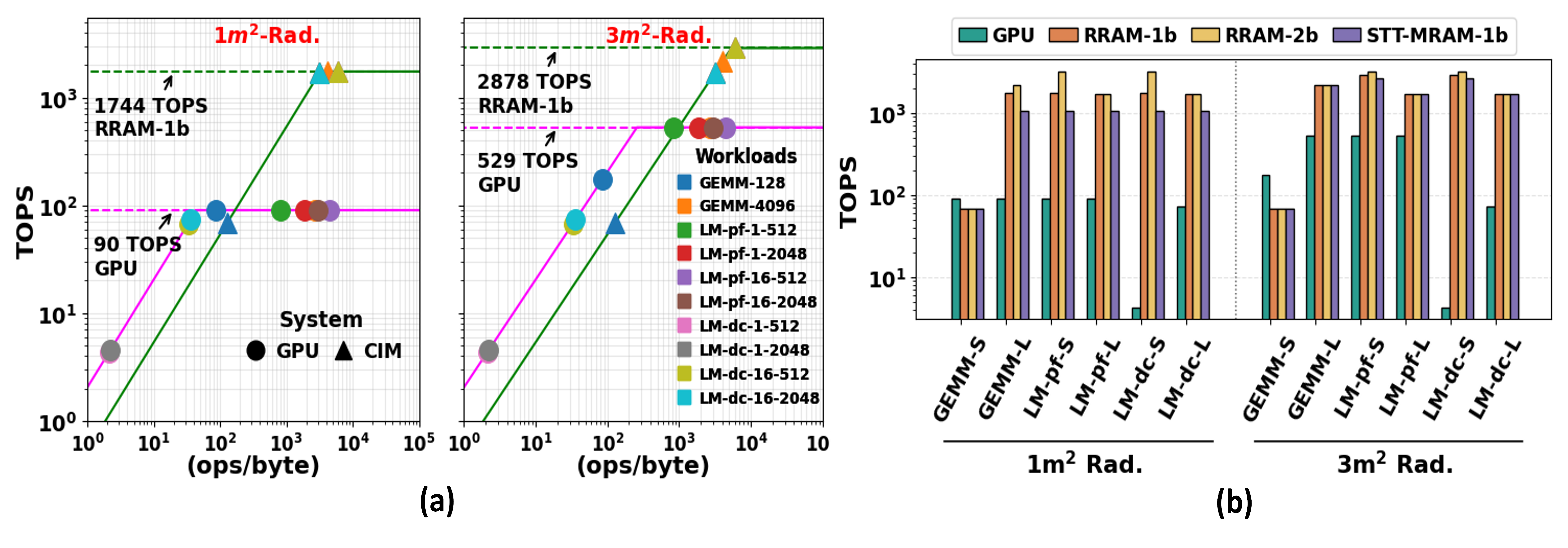}
    \vspace{-20pt}
    \caption{Performance comparison of the GPU and representative CIM systems across workloads under different radiator budgets. (a) shows the roofline TOPS for $1m^2$ radiator (TRP=100 W) and $3m^2$ radiator (TRP=300 W), respectively. (b) summarizes CIM performance advantages over the GPU across workloads and configurations.}
    \vspace{-10pt}
    \label{fig:workload-compare}
\end{figure*}
536 GB/s bandwidth \cite{hbm2}. All evaluations assume 8-bit-precision weights/activations. The experimental system and device parameters are summarized in Table~\ref{tab:system_device}

\textbf{Performance comparison under varying TRP.} We compare three representative CIM configurations with the GPU under varying TRP budgets. These include resistive RAM (RRAM) with 1-bit and 2-bit per cell (R\textsubscript{ON}/R\textsubscript{OFF} = 100k/1M$\Omega$) \cite{puma}, labeled as \textbf{RRAM-1b/2b}, and spin-transfer torque MRAM (STT-MRAM) with 1-bit per cell (R\textsubscript{ON}/R\textsubscript{OFF} = 13k/26k$\Omega$) \cite{mram-ron-roff}, labeled as \textbf{STT-MRAM-1b}. All CIM standard configurations employ 128$\times$128 crossbar arrays with 4-bit ADCs.

As shown in Fig.~\ref{fig:cim-tops-compare}(a), CIM systems consistently achieve higher effective TOPS than the GPU across a wide range of TRP budgets, primarily due to their superior TOPS/W efficiency. All systems cease operation if the TRP constraint forces them below their minimum operating frequency ($f_{min}$). Notably, RRAM-based CIM configurations continue to operate under extremely low power budgets where the GPU must shut down. At the upper end, once systems reach their maximum frequency ($f_{max}$), additional radiator capacity only reduces temperature without improving TOPS. Fig.~\ref{fig:cim-tops-compare}(b) summarizes normalized performance, showing that CIM systems can achieve up to $\sim$40$\times$ higher TOPS than the GPU under stringent radiator constraints (e.g., $1~m^2$ radiator at 100 $W/m^2$). Although the GPU can partially close this gap at higher TRP budgets due to higher achievable clock frequencies, CIM systems maintain overall performance advantages.

\textbf{CIM design exploration.} We evaluate variations in RRAM-based CIM under a constrained TRP budget of 100 W (1 $m^2$ radiator @ 100 $W/m^2$). The explored parameters include the number of array rows (\textbf{Row}), columns per ADC (\textbf{Col}), ADC resolution (\textbf{ADC}), bits per cell (\textbf{BpC}), and on-state resistance (\textbf{ROn}). Each design point varies a single parameter while keeping a baseline configuration of [Row=128, Col=128, ADC=4, BpC=1, ROn=100k$\Omega$]. As shown in Fig.~\ref{fig:cim-design-space}, all CIM configurations outperform the GPU, with the best performance achieved with low-bit ADC and multiple bits/cell. It is important to note that the ADC precision will be bounded by the accuracy requirement and 2-4 bits ADC for 128x128 array have been shown feasible \cite{reconf_adc}. 
%up to $\sim$40$\times$ higher TOPS under low radiator budgets.

\textbf{Workload-level comparison.} Finally, we evaluate representative CIM systems (RRAM-1b/2b and STT-MRAM-1b) and the GPU+HBM system across diverse workloads. Fig.~\ref{fig:workload-compare}(a) presents roofline models for the GPU and a representative CIM configuration (RRAM-1b) under radiator budgets of 1 $m^2$ (left) and 3 $m^2$ (right). The workloads include General Matrix Multiplication between two ($N \times N$) matrices (GEMM-N), where N = 128 and 4096 (\textbf{GEMM-128} and \textbf{GEMM-4096}), as well as Large Language Model (LLM) tasks. The LLM workloads are denoted as \textbf{LM-pf/dc-x-y}, where \textbf{LM} refers to the Llama-3.2-3B model \cite{llama-ref}, \textbf{pf} and \textbf{dc} denote the Prefill and Decode phases, and \textbf{x}, \textbf{y} represent batch size and sequence length, respectively.

The results show that CIM systems achieve higher operational intensity due to weight-stationary dataflow and maintain superior TOPS under radiator constraints, outperforming the GPU across most workloads. Fig.~\ref{fig:workload-compare}(b) summarizes normalized performance across representative workloads, including small (S) and large (L) configurations: \textbf{GEMM-S/L} (128/4096), \textbf{LM-pf/dc-S} (LM-pf/dc-1-512), and \textbf{LM-pf/dc-L} (LM-pf/dc-16-2048). 
Under tight radiator constraints (1 $m^2$), CIM systems significantly outperform the GPU across all workloads. With increased radiator area (3 $m^2$), the GPU reduces the performance gap due to its higher clock frequency and HBM bandwidth, but CIM configurations continue to provide competitive or superior performance.

\section{Conclusion}
\label{sec:conclusion}
We demonstrate the potential of deploying NVM-based CIM accelerators in space under strict radiator-only cooling constraints. We show that conventional GPU+HBM systems suffer from high-power-density hotspots and thermal throttling, constraining performance under limited radiator capacity. In contrast, CIM achieves higher TOPS/W performance thanks to the inherently reduced data movement and uniform power distribution. Our results demonstrate that CIM accelerators consistently deliver over 10-40$\times$ higher effective TOPS than GPU+HBM under tight radiator budgets, while consistently maintaining thermally safe operation across diverse workloads. Our thermal-aware CIM design space exploration and radiator-in-the-loop co-design methodology opens an exciting avenue for developing scalable, energy-efficient AI accelerators for space-based data centers under stringent thermal constraints.
\begin{acks}
This work is supported in part by National Science Foundation (NSF) Grant No. 2441290 and NSF Grant No. 2534279.
\end{acks}
% \setcitestyle{numbers,sort&compress}
\bibliographystyle{ACM-Reference-Format}
\setcitestyle{numbers}
\bibliography{ref}

\end{document}